%% file: main.tex
\pdfoutput=1

\documentclass[12pt,a4paper]{article}

\usepackage{ifthen} 
\newboolean{pdflatex}
\setboolean{pdflatex}{true} 

\newboolean{articletitles}
\setboolean{articletitles}{true} 

\newboolean{uprightparticles}
\setboolean{uprightparticles}{true} 

\newboolean{inbibliography}
\setboolean{inbibliography}{false} 

\input{preamble}
\usepackage{longtable} 

\usepackage{rotating}
\usepackage{wasysym} 
\usepackage{placeins}

\usepackage{graphicx}
\usepackage{caption}

\newcommand{\zmumu}{\ensuremath{\Z\to\mumu}\xspace}

\begin{document}

\renewcommand{\thefootnote}{\fnsymbol{footnote}}
\setcounter{footnote}{1}

\input{title-LHCb-PAPER}


\renewcommand{\thefootnote}{\arabic{footnote}}
\setcounter{footnote}{0}



\pagestyle{plain} 
\setcounter{page}{1}
\pagenumbering{arabic}


%

\input{intro}
\input{detector}

\input{selection}

\input{background}
\input{systematics}
\input{results}
\input{conclusion}
\input{acknowledgements}

\addcontentsline{toc}{section}{References}
\setboolean{inbibliography}{true}
\bibliographystyle{LHCb}
\bibliography{main,local,LHCb-PAPER,LHCb-CONF,LHCb-DP}


\end{document}

%% file: preamble.tex
\usepackage{microtype}
\usepackage{lineno}  
\usepackage{xspace} 
\usepackage{caption} 

\usepackage{graphicx}  
\usepackage{color}
\usepackage{colortbl}
\graphicspath{{./figs/}} 

\usepackage{amsmath} 
\usepackage{amssymb}
\usepackage{amsfonts}
\usepackage{upgreek} 

\newcommand*\patchAmsMathEnvironmentForLineno[1]{%
\expandafter\let\csname old#1\expandafter\endcsname\csname #1\endcsname
\expandafter\let\csname oldend#1\expandafter\endcsname\csname
end#1\endcsname
 \renewenvironment{#1}%
   {\linenomath\csname old#1\endcsname}%
   {\csname oldend#1\endcsname\endlinenomath}%
}
\newcommand*\patchBothAmsMathEnvironmentsForLineno[1]{%
  \patchAmsMathEnvironmentForLineno{#1}%
  \patchAmsMathEnvironmentForLineno{#1*}%
}
\AtBeginDocument{%
\patchBothAmsMathEnvironmentsForLineno{equation}%
\patchBothAmsMathEnvironmentsForLineno{align}%
\patchBothAmsMathEnvironmentsForLineno{flalign}%
\patchBothAmsMathEnvironmentsForLineno{alignat}%
\patchBothAmsMathEnvironmentsForLineno{gather}%
\patchBothAmsMathEnvironmentsForLineno{multline}%
}

\usepackage{hyperref}    
\usepackage[all]{hypcap} 

\input{lhcb-symbols-def} 

\usepackage{cite} 
\usepackage{mciteplus}

%% file: lhcb-symbols-def.tex
\def\lhcb {\mbox{LHCb}\xspace}

\ifthenelse{\boolean{uprightparticles}}%
{
 
 \def\Pgamma      {\ensuremath{\upgamma}\xspace}

 \def\Peta        {\ensuremath{\upeta}\xspace}

 \def\Pmu         {\ensuremath{\upmu}\xspace}

 \def\Ppi         {\ensuremath{\uppi}\xspace}

 \def\PDelta      {\ensuremath{\Delta}\xspace}                 
 \def\PXi      {\ensuremath{\Xi}\xspace}                 
 \def\PLambda      {\ensuremath{\Lambda}\xspace}                 
 \def\PSigma      {\ensuremath{\Sigma}\xspace}                 
 \def\POmega      {\ensuremath{\Omega}\xspace}                 
 \def\PUpsilon      {\ensuremath{\Upsilon}\xspace}                 
 
 \def\PB      {\ensuremath{\mathrm{B}}\xspace}                 
                  
 \def\PD      {\ensuremath{\mathrm{D}}\xspace}

 \def\PK      {\ensuremath{\mathrm{K}}\xspace}

 \def\PW      {\ensuremath{\mathrm{W}}\xspace}

 \def\PZ      {\ensuremath{\mathrm{Z}}\xspace}                 
                  
 \def\Pb      {\ensuremath{\mathrm{b}}\xspace}                 
 \def\Pc      {\ensuremath{\mathrm{c}}\xspace}

 \def\Pi      {\ensuremath{\mathrm{i}}\xspace}

 \def\Pp      {\ensuremath{\mathrm{p}}\xspace}

 \def\Ps      {\ensuremath{\mathrm{s}}\xspace}

}
{
 
 \def\Pgamma      {\ensuremath{\gamma}\xspace}

 \def\Peta        {\ensuremath{\eta}\xspace}

 \def\Pmu         {\ensuremath{\mu}\xspace}

 \def\Ppi         {\ensuremath{\pi}\xspace}

 \mathchardef\PDelta="7101
 \mathchardef\PXi="7104
 \mathchardef\PLambda="7103
 \mathchardef\PSigma="7106
 \mathchardef\POmega="710A
 \mathchardef\PUpsilon="7107
                  
 \def\PB      {\ensuremath{B}\xspace}                 
                  
 \def\PD      {\ensuremath{D}\xspace}

 \def\PK      {\ensuremath{K}\xspace}

 \def\PW      {\ensuremath{W}\xspace}

 \def\PZ      {\ensuremath{Z}\xspace}                 
                  
 \def\Pb      {\ensuremath{b}\xspace}                 
 \def\Pc      {\ensuremath{c}\xspace}

 \def\Pi      {\ensuremath{i}\xspace}

 \def\Pp      {\ensuremath{p}\xspace}

 \def\Ps      {\ensuremath{s}\xspace}

}

\def\mup        {\ensuremath{\Pmu^+}\xspace}
\def\mun        {\ensuremath{\Pmu^-}\xspace} 
\def\mumu       {\ensuremath{\Pmu^+\Pmu^-}\xspace}

\def\W      {\ensuremath{\PW}\xspace}

\def\Wpm    {\ensuremath{\PW^\pm}\xspace}
\def\Z      {\ensuremath{\PZ}\xspace}

\def\squark    {\ensuremath{\Ps}\xspace}

\def\cquark    {\ensuremath{\Pc}\xspace}
\def\cquarkbar {\ensuremath{\overline \cquark}\xspace}
\def\ccbar     {\ensuremath{\cquark\cquarkbar}\xspace}
\def\bquark    {\ensuremath{\Pb}\xspace}

\def\pion  {\ensuremath{\Ppi}\xspace}

\def\pip   {\ensuremath{\pion^+}\xspace}

\def\pipm  {\ensuremath{\pion^\pm}\xspace}

\def\kaon  {\ensuremath{\PK}\xspace}
  \def\Kbar  {\kern 0.2em\overline{\kern -0.2em \PK}{}\xspace}

\def\Km    {\ensuremath{\kaon^-}\xspace}
\def\Kpm   {\ensuremath{\kaon^\pm}\xspace}

  \def\Dbar    {\kern 0.2em\overline{\kern -0.2em \PD}{}\xspace}
\def\D       {\ensuremath{\PD}\xspace}

\def\Dz      {\ensuremath{\D^0}\xspace}

\def\Dp      {\ensuremath{\D^+}\xspace}

\def\Bbar    {\ensuremath{\kern 0.18em\overline{\kern -0.18em \PB}{}}\xspace}

  \def\Y#1S{\ensuremath{\PUpsilon{(#1S)}}\xspace}

\def\proton      {\ensuremath{\Pp}\xspace}

\def\Lbar {\ensuremath{\kern 0.1em\overline{\kern -0.1em\PLambda}}\xspace}

\newcommand{\decay}[2]{\ensuremath{#1\!\to #2}\xspace}         

\def\to                 {\ensuremath{\rightarrow}\xspace}

\def\AT#1     {\ensuremath{A_{\mathrm{T}}^{#1}}\xspace}           

\def\C#1      {\ensuremath{\mathcal{C}_{#1}}\xspace}                       
\def\Cp#1     {\ensuremath{\mathcal{C}_{#1}^{'}}\xspace}                    
\def\Ceff#1   {\ensuremath{\mathcal{C}_{#1}^{\mathrm{(eff)}}}\xspace}        
\def\Cpeff#1  {\ensuremath{\mathcal{C}_{#1}^{'\mathrm{(eff)}}}\xspace}       
\def\Ope#1    {\ensuremath{\mathcal{O}_{#1}}\xspace}                       
\def\Opep#1   {\ensuremath{\mathcal{O}_{#1}^{'}}\xspace}                    

\def\dkpicf     {\decay{\Dz}{\Km\pip}}

\newcommand{\tev}{\ifthenelse{\boolean{inbibliography}}{\ensuremath{~T\kern -0.05em eV}\xspace}{\ensuremath{\mathrm{\,Te\kern -0.1em V}}\xspace}}
\newcommand{\gev}{\ensuremath{\mathrm{\,Ge\kern -0.1em V}}\xspace}
\newcommand{\mev}{\ensuremath{\mathrm{\,Me\kern -0.1em V}}\xspace}
\newcommand{\kev}{\ensuremath{\mathrm{\,ke\kern -0.1em V}}\xspace}
\newcommand{\ev}{\ensuremath{\mathrm{\,e\kern -0.1em V}}\xspace}
\newcommand{\gevc}{\ensuremath{{\mathrm{\,Ge\kern -0.1em V\!/}c}}\xspace}
\newcommand{\mevc}{\ensuremath{{\mathrm{\,Me\kern -0.1em V\!/}c}}\xspace}
\newcommand{\gevcc}{\ensuremath{{\mathrm{\,Ge\kern -0.1em V\!/}c^2}}\xspace}
\newcommand{\gevgevcccc}{\ensuremath{{\mathrm{\,Ge\kern -0.1em V^2\!/}c^4}}\xspace}
\newcommand{\mevcc}{\ensuremath{{\mathrm{\,Me\kern -0.1em V\!/}c^2}}\xspace}

\def\mum  {\ensuremath{{\,\upmu\rm m}}\xspace}

\def\mbarn{\ensuremath{\rm \,mb}\xspace}

\def\pb {\ensuremath{\rm \,pb}\xspace}

\def\invfb   {\ensuremath{\mbox{\,fb}^{-1}}\xspace}

\newcommand{\chisq}{\ensuremath{\chi^2}\xspace}

\def\gsim{{~\raise.15em\hbox{$>$}\kern-.85em
          \lower.35em\hbox{$\sim$}~}\xspace}
\def\lsim{{~\raise.15em\hbox{$<$}\kern-.85em
          \lower.35em\hbox{$\sim$}~}\xspace}

\def\pt         {\mbox{$p_{\rm T}$}\xspace}

\def\fewz       {\mbox{\textsc{Fewz}}\xspace}

\def\tell1  {TELL1\xspace}
\def\ukl1   {UKL1\xspace}



%% file: title-LHCb-PAPER.tex

\begin{titlepage}
\pagenumbering{roman}

\vspace*{-1.5cm}
\centerline{\large EUROPEAN ORGANIZATION FOR NUCLEAR RESEARCH (CERN)}
\vspace*{1.5cm}
\hspace*{-0.5cm}
\begin{tabular*}{\linewidth}{lc@{\extracolsep{\fill}}r}
\ifthenelse{\boolean{pdflatex}}
{\vspace*{-2.7cm}\mbox{\!\!\!\includegraphics[width=.14\textwidth]{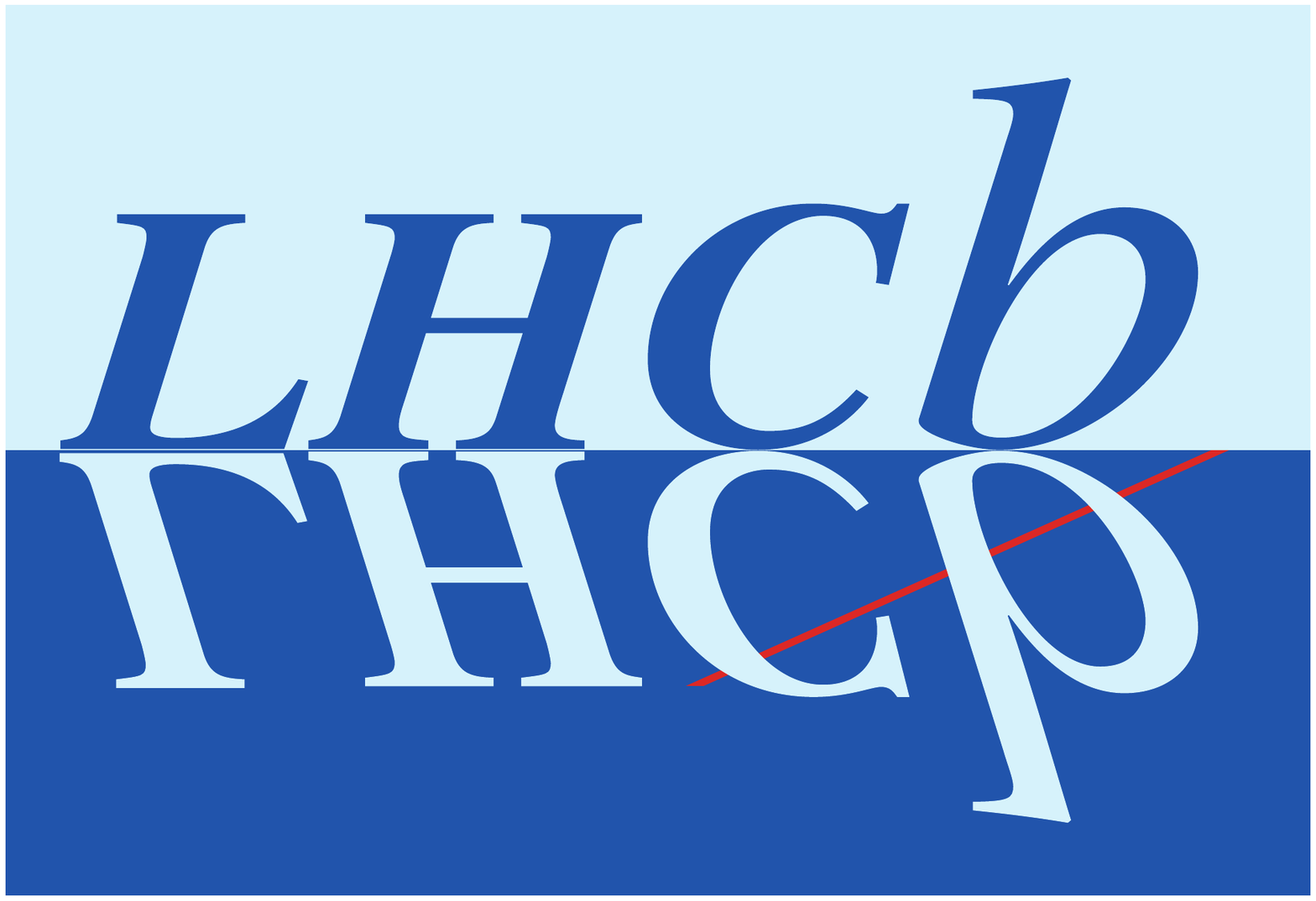}} & &}%
{\vspace*{-1.2cm}\mbox{\!\!\!\includegraphics[width=.12\textwidth]{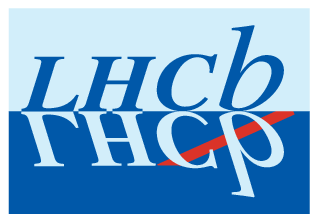}} & &}%
\\
 & & CERN-PH-EP-2013-235 \\  
 & & LHCb-PAPER-2013-062 \\  
 & & \today \\ 
\end{tabular*}

\vspace*{1.5cm}

{\bf\boldmath\huge
\begin{center}
  Observation of associated production of a~\Z~boson with a~\D~meson 
  in the~forward region
\end{center}
}

\vspace*{0.5cm}

\begin{center}
The LHCb collaboration\footnote{Authors are listed on the following pages.}
\end{center}

\vspace{\fill}

\begin{abstract}
  \noindent
  A search for associated production of a~\Z~boson with an~open charm meson 
  is presented using a data sample, corresponding to an integrated luminosity of 
  $1.0\invfb$ of proton--proton collisions at a centre-of-mass energy of  
  7\tev, collected by the~\lhcb experiment.
  Seven candidate events for associated production of a~\Z~boson with
  a~\Dz~meson and four candidate events for a~\Z~boson with a~\Dp~meson 
  are observed with a~combined significance 
  of 5.1~standard deviations.
  The production cross-sections in the~forward region are measured to be
  \begin{eqnarray*}
    \upsigma_{\zmumu\!,\Dz} & = & 2.50\pm1.12\pm0.22\pb \\ 
    \upsigma_{\zmumu\!,\Dp} & = & 0.44\pm0.23\pm0.03\pb,
  \end{eqnarray*}
  where the first uncertainty is statistical and the second systematic.
\end{abstract}

\vspace*{1.0cm}

\begin{center}
  Submitted to JHEP
\end{center}

\vspace{\fill}

{\footnotesize 
\centerline{\copyright~CERN on behalf of the \lhcb collaboration, license \href{http://creativecommons.org/licenses/by/3.0/}{CC-BY-3.0}.}}
\vspace*{2mm}

\end{titlepage}


\newpage
\setcounter{page}{2}
\mbox{~}
\newpage

\input{LHCb_HD_authorlist_2013-11-01}

\cleardoublepage

%% file: LHCb_HD_authorlist_2013-11-01.tex
\centerline{\large\bf LHCb collaboration}
\begin{flushleft}
\small
R.~Aaij$^{40}$, 
B.~Adeva$^{36}$, 
M.~Adinolfi$^{45}$, 
A.~Affolder$^{51}$, 
Z.~Ajaltouni$^{5}$, 
J.~Albrecht$^{9}$, 
F.~Alessio$^{37}$, 
M.~Alexander$^{50}$, 
S.~Ali$^{40}$, 
G.~Alkhazov$^{29}$
% ...

{\footnotesize \it
% ...
$ ^{5}$Clermont Universit\'{e}, Universit\'{e} Blaise Pascal, CNRS/IN2P3, LPC, Clermont-Ferrand, France\\
% ...
$ ^{9}$Fakult\"{a}t Physik, Technische Universit\"{a}t Dortmund, Dortmund, Germany\\
% ...
$ ^{29}$Petersburg Nuclear Physics Institute (PNPI), Gatchina, Russia\\
% ...
$ ^{36}$Universidad de Santiago de Compostela, Santiago de Compostela, Spain\\
$ ^{37}$European Organization for Nuclear Research (CERN), Geneva, Switzerland\\
% ...
$ ^{40}$Nikhef National Institute for Subatomic Physics, Amsterdam, The Netherlands\\
% ...
$ ^{45}$H.H. Wills Physics Laboratory, University of Bristol, Bristol, United Kingdom\\
% ...
$ ^{50}$School of Physics and Astronomy, University of Glasgow, Glasgow, United Kingdom\\
$ ^{51}$Oliver Lodge Laboratory, University of Liverpool, Liverpool, United Kingdom\\
% ...
}
\end{flushleft}

%% file: intro.tex
\section{Introduction}

The forward production cross-section for associated production of 
a~\Z~boson\footnote{The contribution of the virtual $\Pgamma^{*}$  
and charge conjugated modes are always implied in this paper.}
with an~open charm meson
in $\proton\proton$~collisions provides information about
the charm parton distribution inside the proton, the charm production mechanism, 
and  double-parton scattering~\cite{Campbell:2003dd,Campbell:2008cr}.
A~measurement of this cross-section is a~complementary probe to previous measurements by \lhcb of double
charm production~\cite{LHCb-PAPER-2012-003},
inclusive~\Wpm and \Z~boson production~\cite{LHCb-PAPER-2012-008,LHCb-PAPER-2012-036,LHCb-PAPER-2012-029} 
and \Z production in association with jets~\cite{LHCb-PAPER-2013-058}.
Since the~LHCb detector is fully instrumented in the forward region, 
measurements of electroweak boson production at \lhcb have a~unique sensitivity 
to both high and low Bjorken-$x$ regions where parton distribution functions are
not precisely determined by previous measurements~\cite{Thorne:2008am}.

The first observation of associated production of a~\Z~boson with open charm hadrons 
is presented in this paper.
The~ATLAS~and CMS collaborations have recently shown first results of \W~production in association 
with a charmed hadron~\cite{ATLAS-CONF-2013-045,Chatrchyan:2013uja},
a measurement that is directly sensitive to the \squark-quark content 
of the proton.
The~associative production of \Z~bosons with charmed jets 
has been  reported by the~D0~collaboration to be in 
disagreement with next-to-leading order pertubative 
QCD predictions~\cite{Abazov:2013hya}.

In this paper the results are quoted as the~product of the production cross-section and 
the~branching fraction  for the~\zmumu~decay.
The selection of the~\Z~candidates and the~\D~mesons follows those of 
previous publications~\cite{LHCb-PAPER-2012-008,LHCb-PAPER-2012-003,LHCb-PAPER-2013-058},
allowing  the analysis techniques and reconstruction efficiencies to be reused.
The~results are compared to predictions from two production mechanisms:
single-\,(SPS) and double-parton scattering (DPS).

%% file: detector.tex
\section{Detector and data sample}
\label{sec:Detector}

The \lhcb detector~\cite{Alves:2008zz} is a single-arm forward
spectrometer covering the \mbox{pseudorapidity} range $2<\Peta <5$,
designed for the study of particles containing \bquark or \cquark
quarks. The~detector includes a high precision tracking system
consisting of a silicon-strip vertex detector surrounding 
the~$\proton\proton$~interaction region, a~large-area silicon-strip detector located
upstream of a dipole magnet with a bending power of about
$4{\rm\,Tm}$, and three stations of silicon-strip detectors and straw
drift tubes placed downstream.
The combined tracking system provides a momentum measurement with
relative uncertainty that varies from 0.4\% at 5\gev to 0.6\% at 100\gev,
and impact parameter resolution of 20\mum
for tracks with high transverse momentum.\footnote{In this paper units are chosen 
such that $c=1$.}
Charged hadrons are identified
using two ring-imaging Cherenkov detectors~\cite{LHCb-DP-2012-003}. Photon, electron and
hadron candidates are identified by a~calorimeter system consisting of
scintillating-pad and preshower detectors, an~electromagnetic
calorimeter and a~hadronic calorimeter. Muons are identified by 
a~system composed of alternating layers of iron and multiwire
proportional chambers~\cite{LHCb-DP-2012-002}.
The trigger~\cite{LHCb-DP-2012-004} consists of 
a~hardware stage, based on information from the~calorimeter and muon
systems, followed by a~software stage, which applies a~full event
reconstruction.

Candidate events are first required to pass a hardware trigger,
which selects single muons with transverse momentum $\pt>1.48$\gev. In
the subsequent software trigger, at least
one of the final state muons is required to have
$\pt>10$\gev. 
In order to avoid a few events with high hit multiplicity 
dominating the processing time in the software trigger, 
global event cuts are applied.
The dominant global event cut 
requires the total hit multiplicity in the scintillating-pad detector 
to be fewer than 600~hits.
This selects about 90\%~of the events that contain a~\Z~boson.

The data sample consists of~1.0\invfb of integrated luminosity collected 
with the~\lhcb detector in 2011 using $\proton\proton$~collisions at 
a~centre-of-mass energy of 7\tev.

%% file: selection.tex
\section{Event selection}

The selection of \Z~boson candidates and charmed mesons follows those of previous 
publications~\cite{LHCb-PAPER-2012-008,LHCb-PAPER-2012-003,LHCb-PAPER-2013-058}.
Candidate \zmumu~events are selected by requiring a~pair of well 
reconstructed tracks identified as muons. 
The invariant mass of the two muons  must be reconstructed in the range
$60<m_{\mumu}<120\gev$. 
Each muon track must have $\pt>20\gev$ and  
lie in the~pseudorapidity range $2.0<\Peta(\Pmu^{\pm})<4.5$. 
For the reconstruction of $\Dz\to\Km\pip$ and $\Dp\to\Km\pip\pip$ decays,
well reconstructed and identified \pipm and \Kpm candidates are selected.
To ensure a good particle identification separation, the kaons and pions 
are required to be in the momentum range
$3.2 < p < 100\gev$ and $\pt>250\mev$.
The selected hadrons are combined to  form open charm meson candidates  
in the~$\Dz\to\Km\pip$ and
$\Dp\to\Km\pip\pip$ final states in the invariant mass range $1.82<m_{\Km\pip}<1.92\gev$  
for  $\Dz$ and $1.82<m_{\Km\pip\pip}<1.91\gev$ for $\Dp$.
We require $ct$~to be larger than 100\mum, where $t$ is the decay time in the rest 
frame of the open charm mesons.
All open charm mesons are required to have  rapidity reconstructed 
in the range $2<y(\D)<4$ and  $2<\pt(\D)<12\gev$.
The kinematic selection criteria mentioned above, with the exception
of the~requirements on pions and kaons, 
define the fiducial region of this analysis.

The \Z~boson and charmed meson are required to be consistent 
with being produced at the same primary vertex.
This is achieved by a requirement on the~global $\chisq$ of this hypothesis,
which itself is based on the~$\chisq$ of the~impact parameters of the muons and 
the~\D~candidates and the~vertex $\chisq$ of the~reconstructed \D~meson 
candidates~\cite{Hulsbergen:2005pu}.

\begin{figure}[t]
\includegraphics[width=0.5\textwidth]{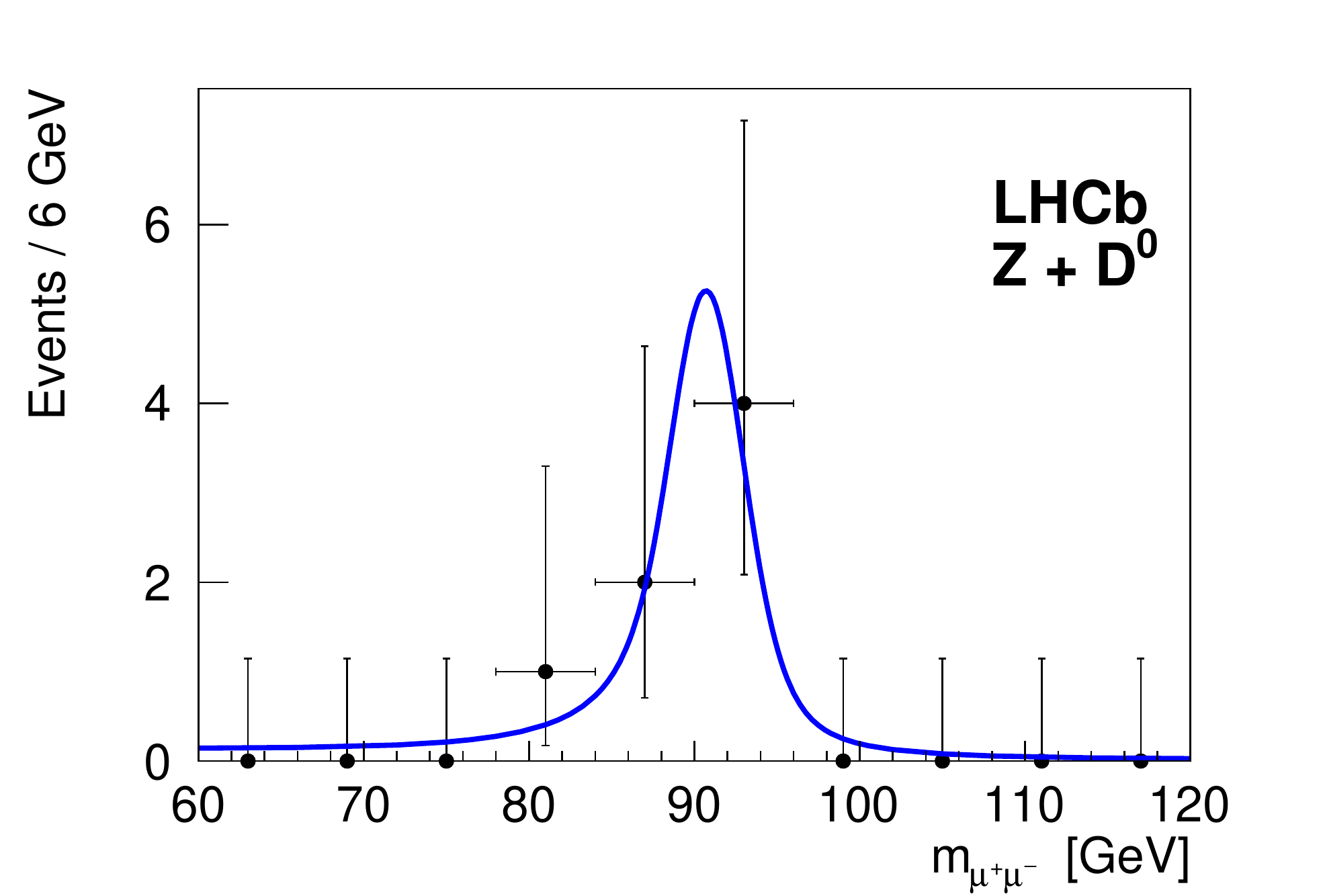}
\includegraphics[width=0.5\textwidth]{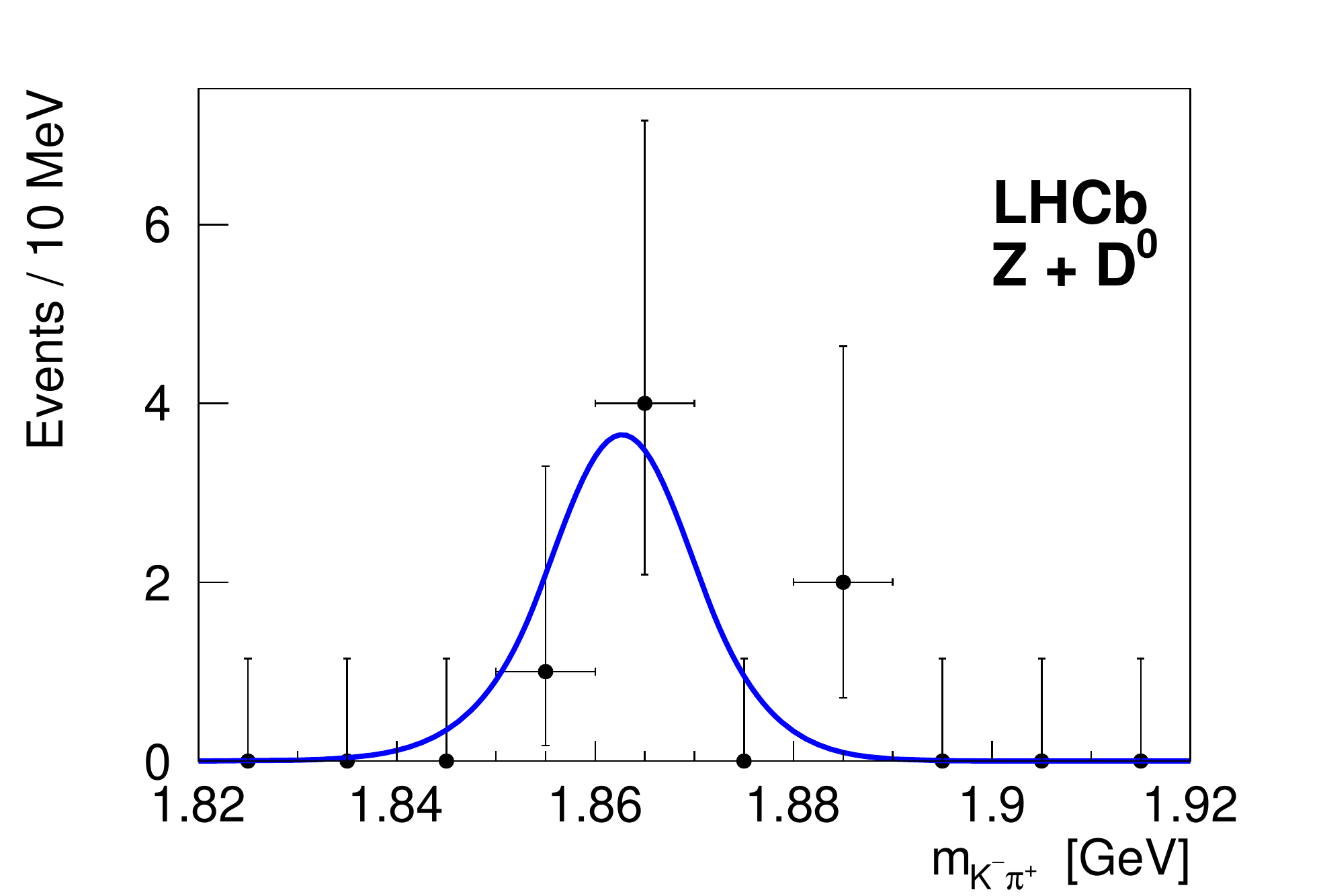}
\includegraphics[width=0.5\textwidth]{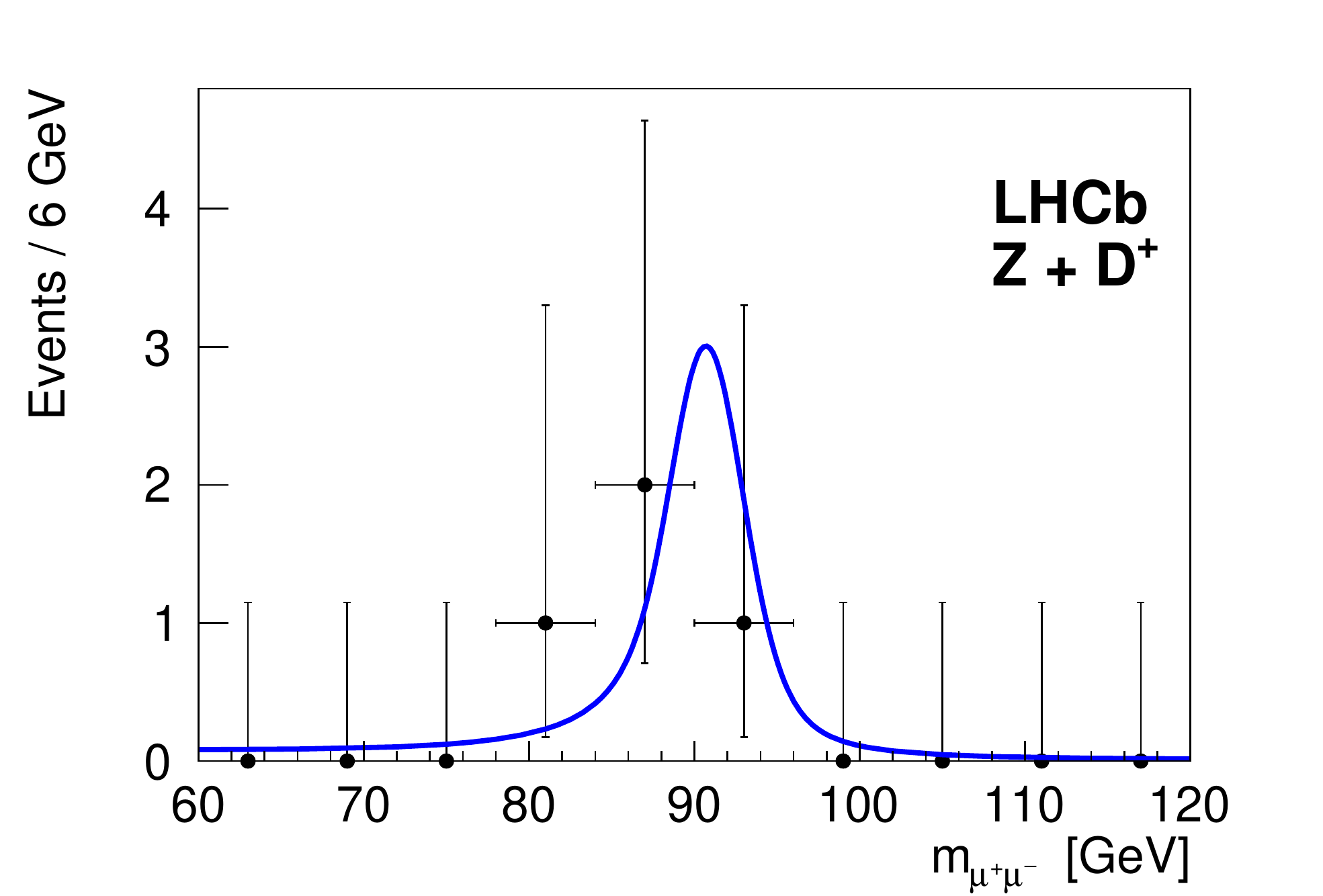}
\includegraphics[width=0.5\textwidth]{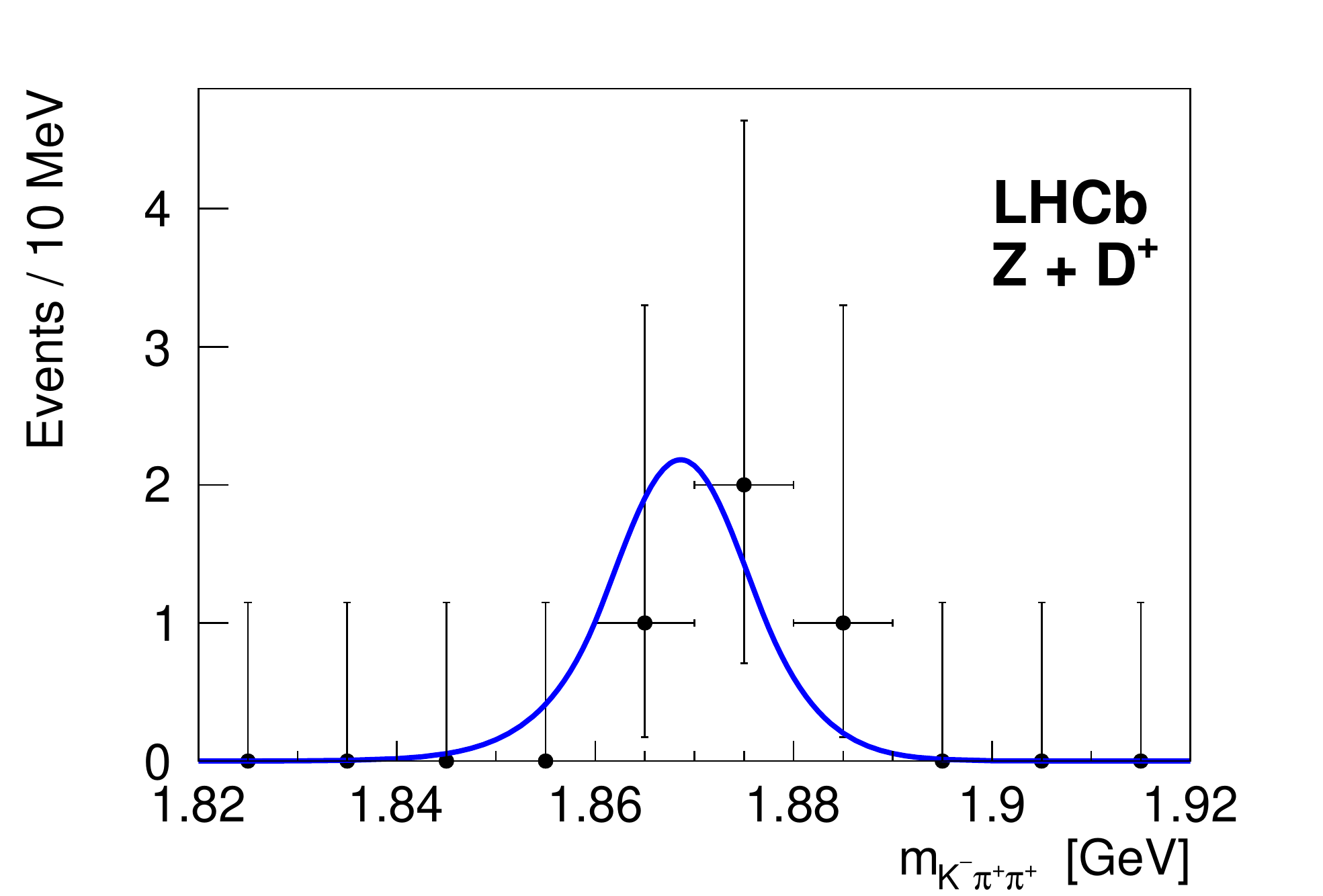}
\caption{\small 
  Invariant mass distribution for \Z(left) and \D(right) candidates
  for $\Z+\Dz$\,(top) and $\Z+\Dp$\,(bottom) events.
  The superimposed curves represent the projection of the fit 
  described in Sect.~\ref{seq:background}.
} 
\label{fig:mass1d}
\end{figure}

In total seven events with \Z~and \Dz~candidates and 
four events with \Z~and \Dp~candidates pass all 
selection criteria, no events with multiple candidates are observed.
The invariant mass distributions for the~\D and the~\Z candidates 
are shown in Fig.~\ref{fig:mass1d}.

%% file: background.tex
\section{Cross-section determination and significance}
\label{seq:background}

Signal events are those for which the Z boson and charmed meson are produced directly 
in  the same $\proton\proton$~interaction.
Charmed hadrons produced from the decay of a beauty hadron are considered as background.
In addition two other background sources are considered: combinatorial background and background 
from multiple $\proton\proton$~interactions~(pile-up).

Both the SPD and DPS mechanisms can lead to the associated production of a~\Z~boson 
and a~beauty hadron. Contamination from feed-down from beauty hadrons decaying to \D~mesons, 
where the beauty hadron has been produced in DPS, is estimated from simulation to be 1.7\%\,(1.3\%) 
for $\Dz(\Dp)$~\cite{LHCb-PAPER-2012-003} of the DPS contribution for a~\Z~boson and a~charmed meson.
The SPS contribution to the feed-down is determined with {\sc{MCFM}}~\cite{Campbell:2010ff},
which predicts the associated production of a~\Z~boson with a~\bquark~quark
to be 20\% smaller than the associated production of a~\Z~with a~\cquark~quark.
This estimate is likely to be conservative, since, 
according to the~recent measurements by the D0~collaboration~\cite{Abazov:2013hya}, 
the production of $\Z+\cquark$-jets  is 
larger by a factor  four with respect to 
$\Z+\bquark$-jets 
for the region with jet $\pt>20\gev$, with only a~small dependence
on the jet $\pt$~\cite{Abazov:2013hya}.
Taking into account the branching fractions, the  beauty feed-down contribution in SPS is estimated 
to be 9.4\%\,(3.7\%) for $\Dz(\Dp)$~mesons of the SPS contribution for a~\Z~boson and a~charmed meson.
This estimate takes into account the suppression due to the requirement on the~\D~to originate from 
the same vertex as the~\Z~candidate.
Since the individual contributions to feed-down from \Z~plus a~\bquark~quark from DPS and SPS are unknown,
we assume that the~contamination from \bquark-quark decays is dominated by DPS.
This assumption is in line with the theoretical predictions for \Z~plus charm quark production 
shown in Table~\ref{tab:xsec}. 
An~uncertainty is assigned that corresponds to the assumption that the SPS contribution is at most 50\%.
This leads to an uncertainty of half the difference between DPS and SPS of 3.9\%\,(1.1\%) for the 
$\Dz(\Dp)$~meson sample.

\begin{figure}[t]
\includegraphics[width=0.5\textwidth]{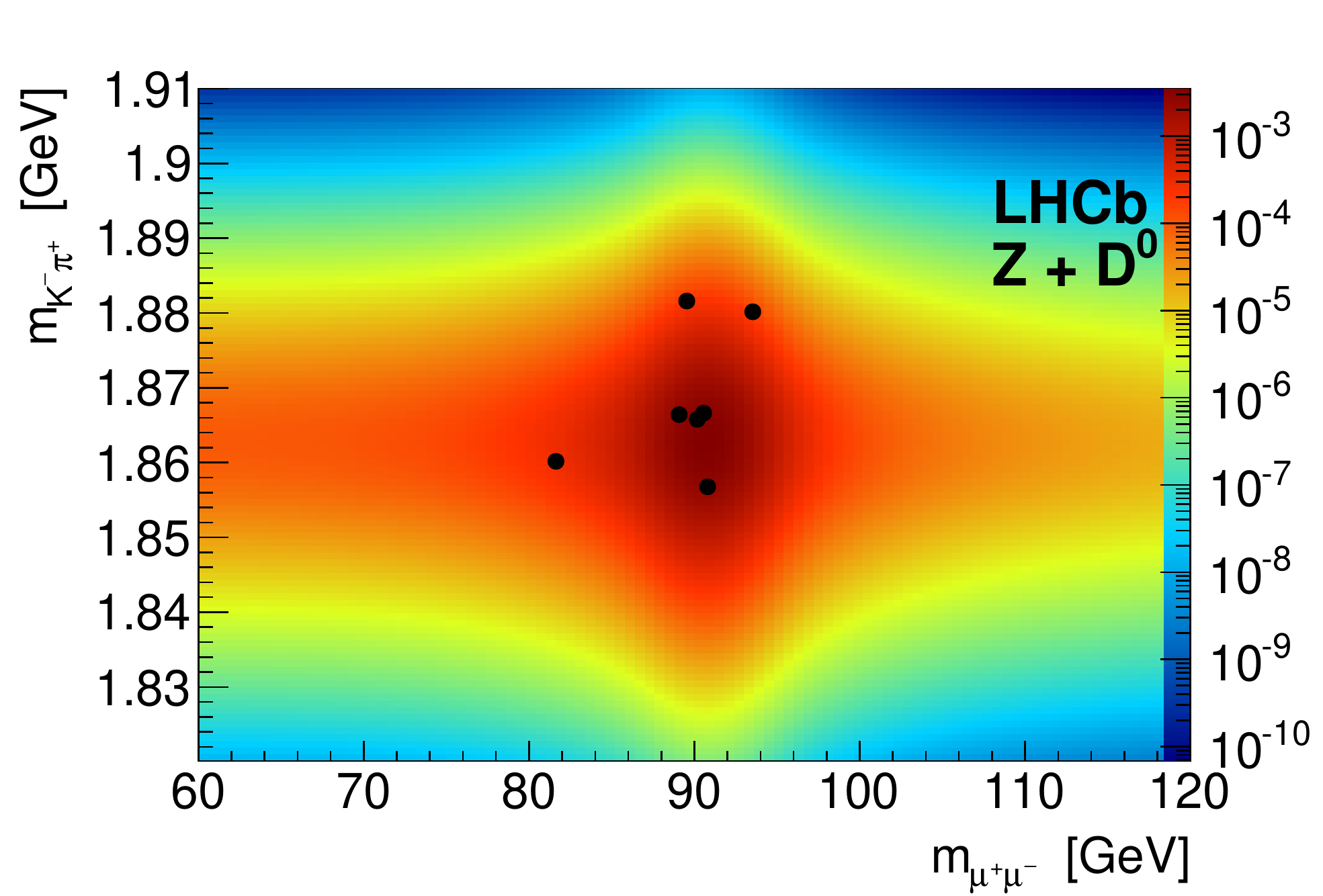}
\includegraphics[width=0.5\textwidth]{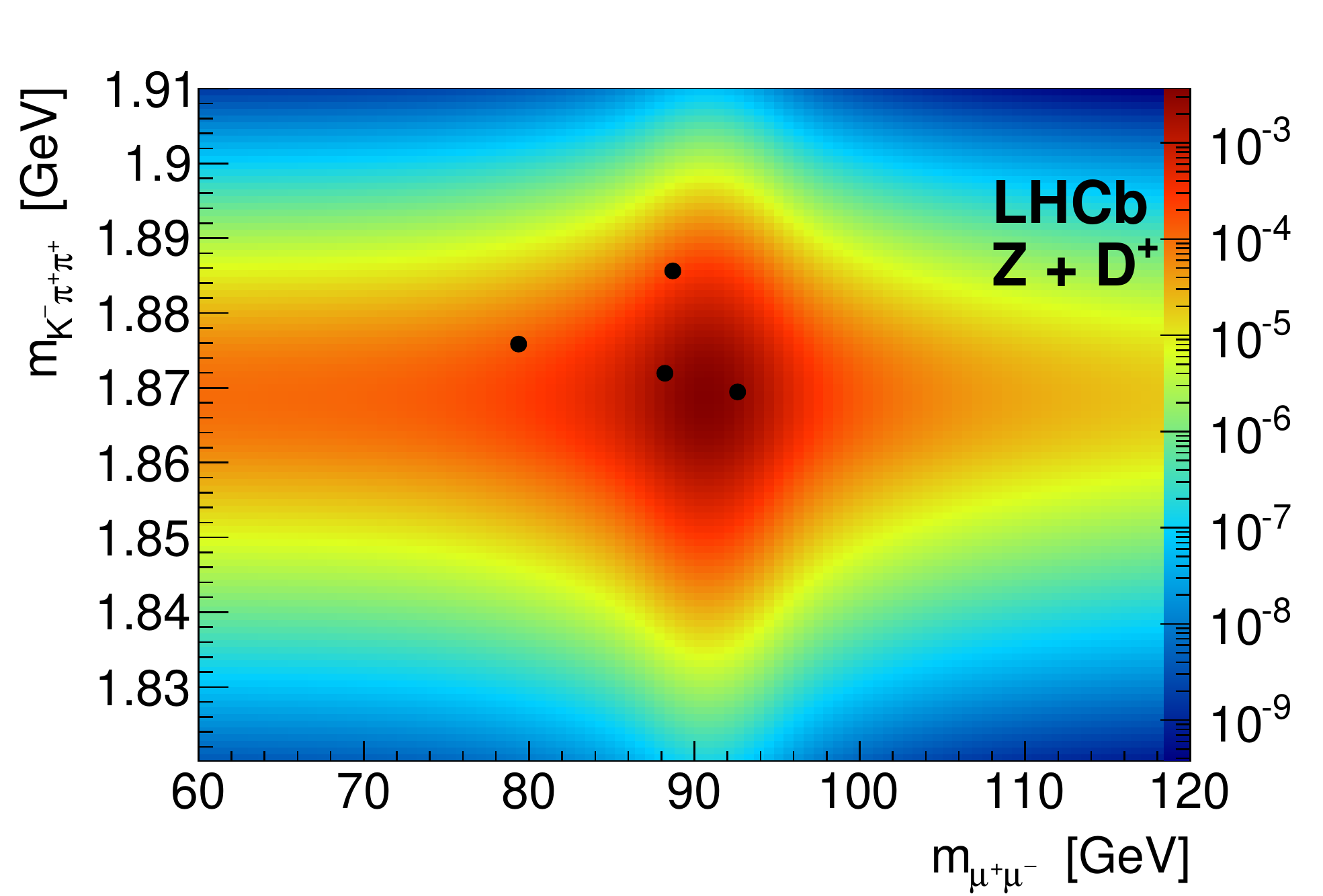}
  \caption{\small
    Invariant mass of the \Z and \Dz (left) and \Z and \Dp (right) candidates (shown as black dots)
    compared to the fit (see text) that was used to extract the combinatorial background. The fit shown includes the signal and the background components. 
    The colour scale shows the PDF value at any given point. 
  }\label{fig:massfit}
\end{figure}

Combinatorial background is estimated by performing a~two-dimensional fit to the  mass distributions  
of the~\Z~boson and the~\D~meson candidates. 
Probability density functions (PDFs) describing the signal and backgrounds are used for the fit: 
 the signal consists of a~\Z~boson with a~\D~meson;
the background consists of a signal \Z~boson with a~random combination of charged hadrons as well as
combinatorial background where all measured stable particles are randomly combined.
Since the combinatorial background for \Z~bosons is known to 
be small $(0.31\pm0.06)\%$~\cite{LHCb-PAPER-2013-058}, 
it is not considered explicitly in the fit model.
The PDF for the \Z~invariant mass is calculated using \fewz~\cite{FEWZ} with the~\Z mass as the renormalisation 
and factorisation scale and using the MSTW08~\cite{Martin:2009iq} parametrisation 
for the~parton density functions of the proton.
Final-state radiation and detector resolution are included by convolving the resulting \Z~lineshape with 
a~resolution function, obtained using the inclusive \Z~sample of the same data taking period.
The~PDF for the charmed hadron candidates is a~modified Novosibirsk function~\cite{Lees:2011gw} 
with the parameters taken from Ref.~\cite{LHCb-PAPER-2012-003}. 
The~combinatorial background components are modelled with exponential distributions 
for the purity determination and a uniform distribution for the significance calculation.
Using a uniform distribution for the combinatorial background in the significance calculation is a conservative approximation:  
it improves the stability of the fit and tends to assign more events to the signal region and therefore leads to a lower significance.
The fit to the two-dimensional mass distributions of the~\Z~boson and the open charm 
candidates is shown in Fig.~\ref{fig:massfit}.

Following Refs.~\cite{LHCb-PAPER-2012-003,Hulsbergen:2005pu},
the contribution from pile-up is assessed using a~fit to the~$\chi^2$~distribution 
of the~hypothesis that the~\Z~boson and the~\D~mesons originate from the same primary vertex.
It is estimated from a~higher statistics sample with a looser selection  to be $(2.8\pm0.6)\%$.
The total purity, defined as the signal fraction, amounts to 
$(95.3\pm3.8)\%$ and $(95.6\pm1.2)\%$ for the \Z~boson plus \Dz~and \Dp~meson samples, respectively. 

The cross-sections are then calculated as 
\begin{equation}
 \upsigma_{\zmumu\!,\D}=
\dfrac{\rho}{\mathcal{L}\,\mathcal{B}_{\D}\, \varepsilon_{\mathrm{GEC}}}\, N^{\mathrm{corr}}_{\zmumu\!,\D} = 
\dfrac{\rho}{\mathcal{L}\,\mathcal{B}_{\D}}\sum_{\text{candidates}}\varepsilon^{-1},
\end{equation}
where
$N^{\mathrm{corr}}_{\zmumu\!,\D}$ is the efficiency-corrected event yield,
$\varepsilon$ is the single event efficiency, $\varepsilon_{\mathrm{GEC}}$ the efficiency of the global event cuts used in the trigger,
$\rho$ the purity, $\mathcal{L}$ the integrated luminosity 
and $\mathcal{B}_{\D}$ the branching fraction of an~open charm hadron into the reconstructed final state~\cite{PDG2012}.

The single event efficiencies are computed according to  
Refs.~\cite{LHCb-PAPER-2012-008,LHCb-PAPER-2012-029,LHCb-PAPER-2013-058,LHCb-PAPER-2012-003} as 
\begin{equation*}
 \varepsilon = \varepsilon^{\mathrm{trg}}_{\zmumu} 
\times \varepsilon_{\zmumu} 
\times \varepsilon_{\D},
\end{equation*}
where $\varepsilon_{\zmumu}$ and  $\varepsilon_{\D}$ are  the~\zmumu~and \D~reconstruction efficiencies, respectively,
and  $\varepsilon^{\mathrm{trg}}_{\zmumu}$ is the trigger efficiency.
The efficiencies  $\varepsilon_{\zmumu}$ and  $\varepsilon_{\D}$ are taken 
from Refs.~\cite{LHCb-PAPER-2013-058} and~\cite{LHCb-PAPER-2012-003}, respectively.
The trigger efficiency $\varepsilon^{\mathrm{trg}}_{\zmumu}$ is calculated as 
\begin{equation*}
\varepsilon^{\mathrm{trg}}_{\zmumu} = 
1 -  
\left( 1 - \varepsilon^{\mathrm{trg}}_{1\Pmu}(\mup) \right) \times 
\left( 1 - \varepsilon^{\mathrm{trg}}_{1\Pmu}(\mun) \right), 
\end{equation*}
where $\varepsilon^{\mathrm{trg}}_{1\Pmu}$ is the efficiency of the~single muon trigger, 
that in turn has been measured using a tag-and-probe 
method on the inclusive \zmumu~sample~\cite{LHCb-PAPER-2012-008}.
All efficiencies have been validated using data-driven techniques 
and the appropriate correction factors  have been applied~\cite{LHCb-DP-2013-001,LHCb-DP-2013-002,LHCb-DP-2012-004,LHCb-DP-2012-003,LHCb-DP-2012-002,LHCb-PAPER-2011-013,LHCb-PAPER-2010-001}. 
The efficiencies have been further corrected for the inefficiency introduced by the global event cuts used in trigger.
Finally, the efficiency corrected yields are found to be 
$N^{\mathrm{corr}}_{\zmumu\!,\Dz}= 99 \pm 45$ and 
$N^{\mathrm{corr}}_{\zmumu\!,\Dp}= 41 \pm 21$,
where the uncertainties are statistical only.

The results of the two-dimensional mass fits described above 
allow the~significance of the~observation 
of the~associated production of a \Z~boson with an~open 
charm meson to be estimated. 
The significance  is assessed using 
pseudo-experiments.
For each pseudo-experiment the events are sampled according to 
the observed number of events using the background-only hypothesis.
The distributions obtained are fitted using the~function described above. 
The~$p$-value obtained from the pseudo-experiments for the associated production 
of \Z~with \D~mesons 
corresponds to a~significance 
of~3.7 and 3.3~standard deviations 
for the~\Dz and \Dp~cases, respectively.
The combined significance for the~associated production of a~\Z~boson 
with an~open charm meson
corresponds to a~significance of 5.1~standard deviations.

%% file: systematics.tex
\section{Systematic uncertainties}

The largest systematic uncertainties are summarised in Table~\ref{tab:systematics}.
The total systematic uncertainties are 8.7\%\,(6.6\%) for the $\Dz(\Dp)$ 
samples and are therefore small with respect to the statistical uncertainties.

Systematic uncertainties on the~trigger, reconstruction and selection efficiencies are computed 
in a~similar manner to Refs.~\cite{LHCb-PAPER-2012-003,LHCb-PAPER-2012-008}. 
 They are dominated by the statistical uncertainty of the tag and probe samples for all 
efficiencies related 
to the~\Z~and differences in the track reconstruction efficiency  between data and 
simulation as well as uncertainties in the particle identification efficiency in case of 
the~\D~reconstruction. 
The uncertainties are propagated by varying the efficiencies ten thousand times within their 
uncertainties 
and taking the standard deviation 
of the~resulting yields as the~uncertainty on the event yield.
In total the estimated uncertainty due to the~efficiencies corresponds to 6.8\%\,(5.0\%) for 
the~$\Dz(\Dp)$~samples.

An uncertainty on the pile-up contamination of 0.6\% is assigned as a~systematic uncertainty.
The feed-down from beauty hadron decays was estimated with  precision of 
3.9\%\,(1.1\%) for \Z~and $\Dz(\Dp)$, and is assigned as a systematic uncertainty.
The~uncertainties in the~branching fractions of an~open charm hadron into the 
reconstructed final state of 1.3\% for \Dz and 2.1\% for \Dp are taken from Ref.~\cite{PDG2012}.

The absolute luminosity scale was measured with a precision of 3.5\,\% at specific periods 
during the~data taking,
using both van der Meer scans~\cite{vandermeer} where colliding beams are moved transversely across
each other to determine the beam profile, and a beam-gas imaging method~\cite{FerroLuzzi:2005em,LHCb-PAPER-2011-015}. 

Other systematic uncertainties, including those related to the~purity estimation are found to be negligible. 

\begin{table}[tb]
  \centering
  \caption{\small
    Relative systematic uncertainties 
    for the production cross-section of a \Z~boson with an~open charm meson 
    $\left[\%\right]$. }
  \label{tab:systematics}
  \begin{tabular*}{0.65\textwidth}{@{\hspace{10mm}}l@{\extracolsep{\fill}}cc@{\hspace{10mm}}}
    & $\Z+\Dz$ & $\Z+\Dp$ \\\hline
    Efficiencies       & 6.8  & 5.0  \\
    Pile-up             & 0.6  & 0.6  \\ 
    Feed down          & 3.9  & 1.1  \\
    $\mathcal{B}_{\D}$  & 1.3  & 2.1  \\ 
    Luminosity         & 3.5  & 3.5  \\ \hline 
    Total              & 8.7  & 6.6    
  \end{tabular*}
\end{table}

%% file: results.tex
\FloatBarrier
\section{Results and discussion}
\label{sec:results}
The cross-sections for associated production of a~\Z~boson and a~\D~meson are measured to be
\begin{eqnarray*}
\upsigma_{\zmumu\!,\Dz} & = & 2.50\pm1.12\pm0.22\pb \\ 
\upsigma_{\zmumu\!,\Dp} & = & 0.44\pm0.23\pm0.03\pb,
\end{eqnarray*}
where the first uncertainty is statistical and the second systematic.
These cross-sections correspond to the following fiducial region: 
\mbox{$60<m_{\mumu}<120\gev$}, 
\mbox{$\pt(\Pmu^{\pm})>20\gev$}, 
\mbox{$2<\Peta(\Pmu^{\pm})<4.5$},
\mbox{$2<\pt(\D)<12\gev$} and 
\mbox{$2<y(\D)<4$}.

The measured cross-section is expected 
to be the sum of the SPS and DPS predictions.
The prediction of the SPS for the~$\Z\ccbar$~production 
cross-section is calculated with MCFM~\cite{Campbell:2010ff} 
at leading order and, using the massless approximation,
at next-to-leading order~\cite{Campbell:2003dd}. 
The~contributions 
from $\Z\cquark$~production~\cite{Campbell:2003hd} are calculated 
in both cases at next-to-leading order.   
The renormalisation and factorisation  scales are set to the~\Z~boson 
mass and varied by a factor of two to assess the theory uncertainty. 
The MSTW08~\cite{Martin:2009iq} parton distribution functions with 
their uncertainties at 68\% confidence level are used.
For the~parton level predictions 
the~fiducial region requirements on the~\D~mesons are applied to 
the~\cquark~quarks. The~cross-sections are corrected 
for the fragmentation fractions as in~Ref.~\cite{LHCb-PAPER-2012-041}.
These hadronisation factors do not take into account  the~change in momentum 
in the $\cquark\to\D$~transition, but only the total probability that 
a~charm quark hadronises into a~given charm meson.
Reference~\cite{Berezhnoy:2012xq} suggests that the~hadronisation 
of charm quarks may lead to an~enhancement of charm hadrons in the \lhcb acceptance.

The DPS cross-section is calculated using
the~factorisation approximation as~\cite{DPS:Stirling}
\begin{equation}
 \upsigma_\text{\zmumu,\D}^{\mathrm{DPS}} = \dfrac{\upsigma_{\zmumu}\,\upsigma_{\PD}}{\upsigma_{\mathrm{eff}}},
 \label{formula:DPS}
\end{equation}
where $\upsigma_{\zmumu}$ and $\upsigma_{\D}$ are the inclusive 
production cross-sections of \zmumu and \D~mesons, respectively,  
and $\upsigma_\text{eff}$~is the~effective DPS 
cross-section.
The production cross-sections of \Z~bosons and  
prompt \D~mesons are taken from Refs.~\cite{LHCb-PAPER-2012-041,LHCb-PAPER-2012-008} 
and extrapolated to the fiducial region of this analysis. 
The~effective DPS cross-section has been measured by several 
experiments at the ISR~\cite{Akesson:1986iv}, 
SPS~\cite{Alitti:1991rd}, Tevatron~\cite{Abe:1997xk,PhysRevD.81.052012} 
and LHC~\cite{Aad:2013bjm,Chatrchyan:2013xxa,LHCb-PAPER-2012-003}.
The~measured value is energy and process independent within 
the experimental precision~\cite{Seymour:2013qka} and 
the~value of $\upsigma_{\mathrm{eff}}=14.5\pm1.7^{+1.7}_{-2.3}\mbarn$ 
is taken  from Ref.~\cite{Abe:1997xk}.
The~factorisation ansatz used to derive Eq.~\eqref{formula:DPS} has 
been criticised as  being too na\"ive~\cite{Blok:2011bu}.   The corresponding uncertainty 
is not assessed here but  could be large in this region  of phase space~\cite{DPS:Stirling}.  
The~contribution of the~non-factorisable component is estimated in 
Ref.~\cite{Korotkikh:2004bz} to be  30\,\% for $x\le0.1$ and 
up to 90\,\% for $x\sim0.2-0.4$.

The measured cross-sections together with three theoretical predictions  are presented 
in Table~\ref{tab:xsec}: a~DPS prediction and two SPS predictions from fixed order calculations 
using {\sc{MCFM}}~\cite{Campbell:2010ff}. 
For the associative production of \Z~bosons and \Dz~mesons 
the~sum of DPS and SPS contributions  is consistent with the measured 
cross-section within the large uncertainties from both theory and experiment, 
while for $\Z+\Dp$~case, the measured cross-section lies below the expectations.

\begin{table}[t]
 \centering
 \caption{\small
         Comparison of the measured cross-sections $\left[\pb\right]$
	 and the theoretical predictions for the~associated production of 
	 a~\Z~boson with an~open charm meson.
	 For the~measured cross-section the~first uncertainty is statistical and 
         the~second systematic. For {\sc{MCFM}} the~first uncertainty 
	 is related to the~uncertainties of the~parton distribution functions, 
         the second is the~scale uncertainty
         and the~third due to uncertainties associated with \cquark-quark 
         hadronisation as discussed in the text. 
        } 
\label{tab:xsec}
\begin{tabular*}{0.99\textwidth}{@{\hspace{2mm}}l@{\extracolsep{\fill}}cccc@{\hspace{2mm}}}  
& measured
& {\sc{MCFM}} massless
& {\sc{MCFM}} massive
& DPS   \\
\hline 
\vspace{-5mm}& & & & \\[10pt]
$\Z+\Dz$~~  &  $2.50\pm1.12\pm0.22$  
         &  $0.85^{+0.12}_{-0.07}       \phantom{|}^{+0.11}_{-0.17}\pm0.05$ 
         &  $0.64^{+0.01}_{-0.01}       \phantom{|}^{+0.08}_{-0.13}\pm0.04$ 
         &  $3.28^{+0.68}_{-0.58}$   \\[10pt]
$\Z+\Dp$  &  $0.44\pm0.23\pm0.03$  
         &  $0.37^{+0.05}_{-0.03}       \phantom{|}^{+0.05}_{-0.07}\pm0.03$ 
         &  $0.28^{+0.01}_{-0.01}       \phantom{|}^{+0.04}_{-0.06}\pm0.02$ 
         &  $1.29^{+0.27}_{-0.23}$    
\end{tabular*}
\end{table}


%% file: conclusion.tex
\section{Conclusion}

Associated production of a~\Z~boson with an~open charm hadron 
is observed by LHCb for the first time in $\proton\proton$~collisions 
at a centre-of-mass energy $\sqrt{s}=7~\tev$ corresponding to 
an~integrated luminosity of $1.0~\text{fb}^{-1}$.

Eleven signal candidates are observed, consisting of seven \dkpicf candidates and four $\Dp\to\Km\pip\pip$ candidates, all associated with a \zmumu decay. 
The cross-sections for the  associated production of \Z~bosons and \D~mesons 
in the fiducial region  are found to be 
\begin{eqnarray*}
\upsigma_{\zmumu\!,\Dz} & = & 2.50\pm1.12\pm0.22\pb \\ 
\upsigma_{\zmumu\!,\Dp} & = & 0.44\pm0.23\pm0.03\pb,
\end{eqnarray*}
where the first uncertainty is statistical and the second systematic.
The results are quoted as  the product of the production  cross-section 
and the branching fraction of the \zmumu~decay.
These cross-sections correspond to the fiducial region 
\mbox{$60<m_{\mumu}<120\gev$}, 
\mbox{$\pt(\Pmu^{\pm})>20\gev$}, 
\mbox{$2<\Peta(\Pmu^{\pm})<4.5$}
\mbox{$2<\pt(\D)<12\gev$} and 
\mbox{$2<y(\D)<4$}.
The results are consistent with the theoretical predictions for $\Z+\Dz$~production, 
and lie below expectations for $\Z+\Dp$~case.
With more data a measurement of the differential distributions will be possible, 
which could allow to disentangle the SPS and DPS contributions.

%% file: acknowledgements.tex
\section*{Acknowledgements}
\noindent 
We thank John~M.~Campbell for help in obtaining the MCFM predictions. 
We express our gratitude to our colleagues in the CERN
accelerator departments for the excellent performance of the LHC. We
thank the technical and administrative staff at the LHCb
institutes. We acknowledge support from CERN and from the national
agencies: CAPES, CNPq, FAPERJ and FINEP (Brazil); NSFC (China);
CNRS/IN2P3 and Region Auvergne (France); BMBF, DFG, HGF and MPG
(Germany); SFI (Ireland); INFN (Italy); FOM and NWO (The Netherlands);
SCSR (Poland); MEN/IFA (Romania); MinES, Rosatom, RFBR and NRC
``Kurchatov Institute'' (Russia); MinECo, XuntaGal and GENCAT (Spain);
SNSF and SER (Switzerland); NAS Ukraine (Ukraine); STFC (United
Kingdom); NSF (USA). We also acknowledge the support received from the
ERC under FP7. The Tier1 computing centres are supported by IN2P3
(France), KIT and BMBF (Germany), INFN (Italy), NWO and SURF (The
Netherlands), PIC (Spain), GridPP (United Kingdom). We are thankful
for the computing resources put at our disposal by
Yandex LLC (Russia), as well as to the communities behind the multiple open
source software packages that we depend on.